\documentstyle[buckow,epsfig,color]{article}

\def\beq{\begin{equation}}
\def\eeq{\end{equation}}
\def\bbbz {{\sf Z\!\!Z}}
\def\Zop{\bbbz}
\def\I{{\cal I}}
\def\dh{$\hat{\mbox{D}}$}
\def\db{$\bar{\mbox{D}}$}
\def\RR{R-R }
\def\NSNS{NS-NS }
\newcommand{\ket}[1]{|#1\rangle}

\begin {document}
\noindent hep-th/0206174 \\
AEI-2002-003 
\def\titleline{D-branes, Orientifolds and K-theory}

\def\authors{
B. Stefa\'nski, jr.\1ad %and Author B \2ad 
}
\def\addresses{
\1ad
Max-Planck-Institut f\"ur Gravitationsphysik,
Albert-Einstein-Institut\\
Am M\"uhlenberg 1,
D-14476 Golm,
Germany \\
{\tt bogdan@aei.mpg.de} 
%\\
%\2ad Address B \\
%Address cont.\\
%{\tt e-mail B}
}
\def\abstracttext{The complete D-brane spectrum in $\Zop_2$ orientifolds 
is computed. Stable non-BPS D-branes with both integral and torsion
charges are found. The relation to K-theory is discussed and a new K-theory 
relevant to orientifolds is suggested.}
\large
\makefront

\section{Introduction}

D-branes~\cite{PolRR} play an important role in string theory. They classify a 
large number of string vacua, provide a canvas for gauge theories and 
are a key ingredent of many dualities. Understanding the spectrum of D-branes
in a given on-shell closed string background is an important and often
difficult subject. For closed string backgrounds with a conformal field theory
description, D-branes are most succinctly represented as boundary 
states~\cite{BCFT} - coherent states constructed out of closed string oscillators.

Another tool used in the study of D-brane spectra is K-theory~\cite{MM,WittK}. 
Since D-branes couple to \RR potentials it was initially thought that they may
be classified by (possibly rational) cohomology. With the discovery of brane
annihilation through tachyon condensation~\cite{Sen} though the K-theory description of
D-branes seems to be the more natural description. The two descriptions differ in 
the classification of torsion-charged branes. Hence, torsion charged D-branes
provide a good test of the K-theory description.

In this note we review the boundary conformal field theory (BCFT) construction 
of D-branes in the $\Omega\times\I_4$ orientifolds~\cite{BiSag,GP,BZDP} of type IIB 
theory~\cite{QS}. We shall focus on D-branes in the GP model and refer the reader 
to~\cite{QS} for details of the BZDP model. The GP model has an O9-plane
and sixteen O5-planes, and in order to cancel the resulting tadpoles one
introduces 16 D9-branes and 16 D5-branes. Placing one D5-brane at each of the
O5-planes one obtains a locally charge canceling configuration. The D5-branes can only
move of the O5-planes in pairs, much as fractional branes in orbifolds. Unlike
fractional branes, they do not couple to twisted sectors and have thus been dubbed
stuck~\cite{QS}. 

When classifying D-branes, we will find it convenient to
follow the notation of~\cite{GS}: we will refer to a D$p$-brane as a D$(r,s)$-brane 
with $r+s=p$ and the brane having $r+1$ Neumann direction in the non-compact 
six-dimensional spacetime and $s$ Neumann directions on $T^4$. For example the tadpole
canceling D5- and D9-branes will be labeled as D$(5,0)$- and D$(5,4)$-branes, respectively.

The rest of this note is organised as follows. In section~\ref{sec2} we discuss integrally
charged non-BPS D-branes and their decay channels, while in section~\ref{sec3} we
present some examples of torsion charged non-BPS D-branes. In section~\ref{sec4} we
compare the BCFT construction to K-theory predictions; we show that the BZDP model
is described by orthogonal equivariant K-theory. In particular there is good agreement 
between K-theory and BCFT for torsion charged branes. We also 
suggest a new twisted version of the equivariant orthogonal K-theory which should 
describe D-branes in the GP orientifold.

\section{Integrally charged non-BPS D-branes}\label{sec2}

Integrally charged non-BPS D-branes carry twisted \RR charges  but are neutral under the 
untwisted \RR charge. In the GP model two types of such truncated branes were encountered.
Firstly $\Omega$-invariant truncated branes from the $\I_4$ orbifold~\cite{GS} are present for
$r=-1,3$ and $s=1,3$. Secondly, D\db pairs, in which $\Omega$ maps the D-brane to the 
\db-brane, occur for $r=-1,3$ and $s=0,4$. In the orbifold the latter are bound 
states of two fundamental objects; in the orientifold however, they form one object.

The decay channels and stability regions of the $\Omega$-invariant truncated branes are 
modified by the $\Omega$ projection. In the orbifold these are stable 
for\footnote{We take $\alpha^\prime=1$ throughout.}
\begin{eqnarray}
R_\perp&\ge&\frac{1}{\sqrt{2}}\,,\label{eq1}\\
R_{||}&\le&\sqrt{2}\label{eq2}
\end{eqnarray}
where $R_\perp$, $R_{||}$ are radii of the compact directions inverted by $\I_4$ 
which are perpendicular and parallel to the truncated brane. For small $R_\perp$ the 
\dh$(r,s)$-brane decays into a D\db pair of superimposed $(r,s+1)$-branes, while for
large $R_{||}$ it decays into a  D\db pair of $(r,s-1)$-branes at opposite fixed points.
In the orientifold the $s=1$ ($s=3$) \dh-branes decay along $x^\perp$ ($x^{||}$) as in 
the orbifold.\footnote{This is consistent since there are $r=-1,3$, $s=2$ BPS fractional
branes in the GP model.} On the other hand the orientifold $s=1$ ($s=3$) \dh-branes 
are stable for all values of $R_{||}$ ($R_\perp$) as the $\Omega$ projection removes the
open-string momentum (winding) states.\footnote{This is fortunate as in the GP orientifold
there are no BPS fractional $r=-1,3$ $s=0,4$ branes into which the truncated branes 
could decay.}

\section{Torsion charged non-BPS D-branes}\label{sec3}

In the GP model two kinds of torsion charged D-branes were encountered. The first type,
present for $s=2$, couples to both the untwisted and twisted \NSNS sector
\begin{equation}
\ket{h_\pm,(r,s=2)}=\ket{NSNS}\pm\ket{NSNS,T}\,,\label{hbrane}
\end{equation}
and in the non-compact theory carry $\Zop_2\oplus\Zop_2$ charge. The second kind 
of torsion-charged branes couple only to the untwisted \NSNS sector
\begin{equation}
\ket{g,(r,s)}=\ket{NSNS}\,.\label{g}
\end{equation}
The latter branes are very similar to the torsion branes encountered in $\I_n\Omega$ 
orientifolds of type II theories~\cite{Sen,WittK,Frau}. We focus here on $r=5$ branes 
and refer the reader to~\cite{QS} for other cases. 

In the uncompactified theory there are two (5,2)-branes with 
untwisted and twisted \NSNS couplings: $h_+$ and $h_-$, both of which are 
stable for\footnote{In particular they are stable in the decompactified theory.}
\begin{equation}
\frac{1}{R^2_{||}}+R_\perp^2\ge\frac{1}{2}\,.
\end{equation}
Further, it was shown in~\cite{QS} that
\begin{equation}
\ket{h_+,(5,2)}+\ket{h_-,(5,2)}=\ket{g(5,2)}\,.
\end{equation}
$g$ can be thought of as a $Z_2$ D7-brane of type I and its image under $\I_4$ and so
satisfies
\begin{equation}
\ket{g}+\ket{g}=\ket{0}\,,
\end{equation}
with $\ket{0}$ representing the closed string vacuum. Since $h_+$ and $h_-$ are distinct, 
the (5,2)-branes in the non-compact theory carry $\Zop_2\oplus\Zop_2$ charge. In the 
compact theory the $h$-branes will couple to four twisted \NSNS sectors, and by turning on 
suitable Wilson lines and fluxes in the compact directions one finds many more 
$h$-branes~\cite{QS}. 

The above torsion charged D7-branes are in fact unstable since the open
string stretching between them and tadpole-canceling D-branes has a tachyonic groundstate.
The D7-branes will as a result decay; the
above discussion does however indicates the presence of a torsion charge in the theory. 
In the BZDP orientifold we found~\cite{QS} torsion-charged D-branes (for example 
a D-particle) which have no such tachyonic instabilities. It is interesting to 
note that such a D-particle is {\em stable} for all values of compactification radii.

The $g$ D7-brane of equation~(\ref{g}), has no tachyonic states from open strings with
both endpoints on its worldvolume for radii satisfying 
equations~(\ref{eq1}) and~(\ref{eq2}). For radii outside these regions 
the D7-brane becomes unstable and decays. The decay channels
are most easily found by considering the brane as an $\Omega$-invariant
superposition of fractional (5,2)-branes in the $\I_4$ orbifold as depicted in 
figure~\ref{fig1}. The decays in the orbifold are known and turn
out to be $\Omega$-invariant. For example for $R_{||}\ge\sqrt{2}$ the g(5,2)-brane
decays into a g(5,1)-brane (also constructed in~\cite{QS}) 
which carries $\Zop_2$ charge. This D6-brane for $R_{||}\ge\sqrt{2}$ 
decays into a stuck D(5,0)-brane at one fixed point and stuck \db(5,0)-brane at another.
The \db(5,0)-brane can annihilate with a tadpole canceling D(5,0)-brane, should the latter be present
at that fixed point. In this case the configuration ends up being BPS and further changes of radii 
will not affect it. On the other hand if there is no tadpole canceling D(5,0)-brane at the
fixed point the configuration remains non-BPS.

\begin{figure}[htb]
\begin{center}
\input{fig5.pstex_t}
\end{center}
\caption{{\scriptsize 
The decay channels of $\Zop_2$-charged branes are most easily seen as an $\Omega$ invariant
process in the $\I_4$ orbifold. The first line in the figure shows the standard decent of an 
$s=2$ D\db pair (a), via an $s=1$ \dh-brane (b), into an $s=0$ D\db pair. The second line is the
$\Omega$-image of this decay. Together, the diagrams show the decays between 
$\Zop_2$-charged $s=2,1,0$-branes in the GP orientifold. 
{\bf (a)} A $\Zop_2$-charged (5,2)-brane (called $h$ in the text) is an $\Omega$ 
invariant configuration 
of four fractional $(5,2)$-branes in the orbifold. The twisted \RR charges of each of the fractional
branes are shown as $\pm $ next to the fixed points denoted by crosses. The untwisted \RR charge is 
$\pm 1$ and shown in the middle of each brane. 
{\bf (b)} A $\Zop_2$ charged (5,1)-brane is 
an $\Omega$ invariant configuration 
of two truncated $(5,1)$-branes in the orbifold. The twisted \RR charges of 
each of the \dh-branes are 
shown as $\pm $ next to the fixed points denoted by crosses. 
{\bf (c)} In the orbifold cover a stuck (5,0)-(anti-)brane is an $\Omega$-invariant 
pair of fractional 
$(5,0)$-(anti-)branes with opposite twisted \RR charges. Here we show 
a stuck brane at one fixed point with 
a stuck anti-brane at the other.}}
\label{fig1}
\end{figure}

Unlike the original D7-brane, both final configurations have no tachyons coming from
open strings with one endpoint on the torsion-charged configurations and one on a tadpole
canceling D-brane; they describe the endpoint of condensation of these tachyons. We 
note that the positions of D5-branes are T-dual to Wilson lines on D9-branes and 
so the $g$ D7-branes can also be thought to decay into gauge configurations on the 
worldvolumes of the D9-branes.

\section{Orientifolds and K-theory}\label{sec4}

D-branes on orbifolds are described by equivariant K-theory. On the other hand,
D-branes in Type II theories projected by $\I_n\Omega$ are classified by Real K-theory. 
It is natural then to expect that 
D-branes in orientifolds of the form $\Omega\I_n\times G$ be classified by Real
$G$-equivariant K-theory. It is in fact straightforward to compute\footnote{By $KR_{\Zop_2\times\Zop_2}$
we mean K-theory with an anti-linear involution $\tau$ and a linear order two map $g$
which together form the $\Zop_2\times\Zop_2$ group.}
\begin{equation}
KR_{\Zop_2\times\Zop_2}(pt)=\Zop\oplus\Zop\,.
\end{equation}
As a result D9-branes classified by $KR_{\Zop_2\times\Zop_2}$ should carry two 
integral charges - untwisted and twisted \RR charges. The D(5,4)-branes of GP
only carry untwisted \RR charge\footnote{This is immediate since there is no
twisted sector tadpole in this model.} and hence  $KR_{\Zop_2\times\Zop_2}$ {\em cannot}
describe this model. Nonetheless the D(5,4)-branes of BZDP do carry both
untwisted and twisted \RR charges\footnote{In the BZDP
model the twisted tadpole has to be canceled.}. Hence, $KR_{\Zop_2\times\Zop_2}$ can only
describe D-branes in the BZDP model. As further evidence of this using long
exact sequences much like those in~\cite{GS} one can for example show that
\begin{eqnarray}
KR_{Z_2\times\Zop_2}(R^{0,1})&=&\Zop_2\oplus\Zop_2\,,\\
KR_{Z_2\times\Zop_2}(R^{1,0})&=&\Zop\,,
\end{eqnarray}
indicating that there should be $\Zop_2\oplus\Zop_2$ charged $(4,4)$-branes
and $\Zop$ charged $(5,3)$-branes in the BZDP model. 
Such D-branes have been found to be consistent using BCFT techniques~\cite{QS}.

D-branes in the GP orientifold are not described by $KR_{\Zop_2\times\Zop_2}$, and at first sight
there is no obvious alternate candidate for it. A similar
problem occurs in $\Zop_2\times\Zop_2$ orbifolds of Type II theories. There
are in fact two such orbifolds~\cite{vafadt} which differ by the action of $g_i$,
on the $g_j$-th twisted sector ($i\neq j$). Since the closed string spectrum differs in 
the two theories, the stable D-branes are also distinct~\cite{z2torsion}. In fact 
D-branes in the theory with no torsion are described by the equivariant K-theory
$K_{\Zop_2\times\Zop_2}$ while those in the theory with torsion by 
$K^{[c]}_{\Zop_2\times\Zop_2}$, a twisted equivariant K-theory.\footnote{$c$ is
related to the non-trivial element of $H^2(\Zop_2\times\Zop_2,U(1))=Z_2$.}
Bundles classified by $K^{[c]}_{\Zop_2\times\Zop_2}$ differ from those of
$K_{\Zop_2\times\Zop_2}$  in that the two generators of $\Zop_2\times\Zop_2$
anticommute as linear maps on the fibres in the twisted case. The GP and BZDP 
models differ by the action of $\Omega$ on the $\I_4$-twisted sector, and so
D-branes in the GP model should be described by $KR^{[c]}_{\Zop_2\times\Zop_2}$: a twisted
version of $KR_{\Zop_2\times\Zop_2}$. 
Bundles classified by $KR^{[c]}_{\Zop_2\times\Zop_2}$ will have anti-commuting
complex conjugation and $\Zop_2$ action on the fibres. Such a K-group
has not been previously studied and it is not clear whether it forms part of
some generalised cohomology theory. For a detailed discussion of these issues see~\cite{BS}.

\section{Conclusions}\label{sec5} 

We have discussed the full spectrum of D-branes in $\I_4\times\Omega$ orientifolds.
In particular non-BPS D-branes with torsion and integer charges were identified,
and their stability regions and decay channels were discussed. The BCFT constructions
were compared to K-theory; we saw in particular good agreement between BCFT and
K-theory for torsion-charged D-branes, supporting the K-theory conjecture. We have
also found that Orthogonal (or Real) equivariant K-theory does not always 
describe the spectrum of D-branes in orientifolds. Rather, there are twisted
versions of such K-groups which are also relevant.

It would be interesting to find if the twisted K-groups form a generalised
cohomology. We hope to comment on this in the near future~\cite{BS}. The torsion
charges we have identified should have an interpretation in terms of dual heterotic
and F- theories. Understanding these charges in the dual theories would provide a 
very non-trivial, non-BPS test of the conjectured dualities.

\section*{Acknowledgments}

The author would like to thank the organisers of the Corfu Summer Institute on Elementary 
Particle Physics for financial support and a stimulating programme. This work was in part
supported by the European Commission RTN program HPRN-CT-2000-00131.

%%%%%%%%%%%%%%%%%%%%%%
%%%%%%%%%%%%%%%%%%%%%%

\end{document}